\documentclass[preprint]{aastex}
\usepackage{graphicx}
\usepackage{txfonts}
\usepackage{url}

\slugcomment{Submitted to ApJ on July 19, 2011. Accepted on March 23, 2012.}

\shorttitle{Spectroscopic signature of Alfv\'en waves damping in a coronal hole}
\shortauthors{Bemporad \& Abbo}

\begin{document}

\title{Spectroscopic signature of Alfv\'en waves damping\\
in a polar coronal hole up to 0.4 solar radii}

\author{A. Bemporad \& L. Abbo}
\affil{Istituto Nazionale di Astrofisica (INAF), Osservatorio Astrofisico di Torino, \\ Via Osservatorio 20, 10025 Pino Torinese, Torino, Italy} 
\email{bemporad@oato.inaf.it}

\begin{abstract}
Between February 24-25, 2009, the EIS spectrometer onboard the Hinode spacecraft performed special ``sit \& stare'' observations above the South polar coronal hole continuously over more than 22 hours. Spectra were acquired with the 1'' slit placed off-limb covering altitudes up to 0.48 R$_\odot$ ($3.34\times 10^2$ Mm) above the Sun surface, in order to study with EIS the non-thermal spectral line broadenings. Spectral lines such as Fe~{\sc xii} $\lambda$186.88, Fe~{\sc xii} $\lambda$193.51, Fe~{\sc xii} $\lambda$195.12 and Fe~{\sc xiii} $\lambda$202.04 are observed with good statistics up to high altitudes and they have been analyzed in this study. Results show that the FWHM of Fe~{\sc xii} $\lambda$195.12 line increases up to $\simeq 0.14$ R$_\odot$, then decreases higher up. EIS stray light has been estimated and removed. Derived electron density and non-thermal velocity profiles have been used to estimate the total energy flux transported by Alfv\'en waves off-limb in polar coronal hole up to $\simeq 0.4$ R$_\odot$. The computed Alfv\'en wave energy flux density $f_w$ progressively decays with altitude from $f_w \simeq 1.2 \cdot 10^6$ erg cm$^{-2}$ s$^{-1}$ at 0.03 R$_\odot$ down to $f_w \simeq 8.5 \cdot 10^3$ erg cm$^{-2}$ s$^{-1}$ at 0.4 R$_\odot$, with an average energy decay rate $\Delta f_w / \Delta h \simeq -4.5 \cdot 10^{-5}$ erg cm$^{-3}$ s$^{-1}$. Hence, this result suggests energy deposition by Alfv\'en waves in a polar coronal hole, thus providing a significant source for coronal heating.
\end{abstract}


\section{Introduction}

Many different physical processes have been proposed to explain the heating and acceleration of plasma emanating from polar coronal holes and resulting in the fast solar wind streams. Basically two competing theories have been proposed: the first one envisages ubiquitous magnetic reconnections occurring at the base of the corona \citep[the so-called nano-flare heating, originally proposed by][]{parker1983,parker1988}, leading to a dissipation of magnetic stresses produced by fieldlines footpoints motions at the coronal base (also referred to as direct current heating). The second theory invoques the propagation and dissipation along magnetic field lines of different types of MHD waves (e.g. magnetoacoustic, Alfv\'en, ion-cyclotron waves), also referred to as alternating current heating \citep[see e.g.][for a review on the coronal heating problem]{klimchuk2006}. More recently, \citet{depontieu2011} have suggested that also a ubiquitous coronal mass supply in which chromospheric plasma (i.e. spicules) is accelerated upward into the corona, and they provide constraints on the coronal heating mechanisms. For this reason, over the last decades many attempts have been performed to provide evidence for detection of waves in the corona and to quantify the energy provided by wave dissipation. A complete review of wave detections in the corona is far from the purposes of this work: here we briefly review main works published on the detection of waves as a broadening of spectral line profiles observed in corona above the limb. This effect, which has been observed with spectroscopic, spectropolarimetric and interferometric measurements, is often interpreted as a signature of Alfv\'en waves propagating undamped in a density stratified corona, because these waves are expected to be responsible for unresolved plasma motions occurring also with a component along the line of sight (both towards and away from the observer), hence leading to a spectral line profile broadening. If this interpretation is correct, a detection of a decrease, instead of an increase, of the line width above a given altitude, would imply a damping of Alfv\'en waves, hence energy deposition in the corona.

One of the first evidences of increasing line profile widths with altitude was obtained during a total solar eclipse by \citet{chandra1991} with a Fabry-Perot (FP) interferometer: they found a rise of the Fe~{\sc xiv} $\lambda 5303$ (\AA) coronal line-width from 0.07 R$_\odot$ (measured from the solar limb) to 0.18-0.2 R$_\odot$ and then a tendency to fall off with higher radial distance. More recently, this result was confirmed by \citet{mierla2008}, who studied the variation in the Fe~{\sc xiv} line width in the inner corona (above 0.1 R$_\odot$) with LASCO/C1 (1996-1998) data, acquired with a FP interferometer used as a narrow-passband, tunable filter. \citet{mierla2008} found that the Fe~{\sc xiv} line width was roughly constant or increased with height up to around 0.3 R$_\odot$, and then it decreased. Even if the interpretation of these line broadenings as signatures of Alfv\'en waves is not universally accepted, recent results demonstrate that the lower corona is permeated by these waves. \citet{tomczyk2007} identified low-frequency ($<$5mHz) propagating Alfv\'enic motions in the solar corona from linear polarization images of the Fe~{\sc xiii} $\lambda 1075$ line. In the chromosphere Alfv\'enic motions have been observed by \citet{depontieu2007a} with Hinode/SOT, and \citet{depontieu2007b} have studied their relationship with chromospheric spicules observed at solar limb. Moreover, it has been demonstrated by Hinode/XRT data that coronal X-ray polar jets occurs at a larger rate than previously reported ($\sim 10$ events per hour) and that these jets are associated with small scale mass ejections propagating at velocities close to what expected for Alfv\'en waves \citep{cirtain2007}.

A very large number of published work were based on spectroscopic observations. First studies date back to the Skylab data and have shown an increase of line width as a function of height above the limb \citep[lines emitted by chromospheric or transition region ions with T$<2.2 10^5$ K;][]{doschek1977, nicolas1977, mariska1978}. \citet{hassler1990} found an increase in the line width up to 0.2 R$_\odot$ above the limb by analyzing several line profiles from the off-limb quiet corona. Studies of the equatorial coronal regions from SOHO data (SUMER and CDS) have confirmed the broadening of spectral line width as a function of height above the limb \citep[e.g.][]{wilhelm2004, harrison2002, doyle1998}. In polar regions, \citet{banerjee1998,doyle1999}, analyzing Si~{\sc viii} spectra by SUMER, have found an increase in the non-thermal velocity up to 0.25 R$_\odot$. This result was confirmed also by \citet{wilhelm2004} through Mg~{\sc x} line observations provided by SUMER in a polar coronal hole. \citet{oshea2003} observed, from spectra obtained by CDS, a decrease of the Mg~{\sc x} line width at 0.15 R$_\odot$, where the dominant excitation changes from collisional to radiative.

While there is a general consensus that EUV line profiles broden in polar coronal holes moving off-limb up to $\sim 0.15-0.20$ R$_\odot$, the line width behaviour above that altitude is not well determined. For instance, some authors found that the line width profile shows a ``plateau'' above a given altitude \citep[e.g.][]{contesse2004,wilhelm2005}, instead of a decrease. Moreover, \citet{dollasolomon2008} reported detection of Alfv\'en waves from the study of the variation of line widths in coronal hole off-limb region by analyzing SUMER/SOHO data, but they claim no evidence for damping of the Alfv\'en waves because they found that the effect of stray light explained the observed decrease with height in the width of several spectral lines, starting about 0.1--0.2 R$_\odot$ above the limb. This result was also confirmed from a later study by \citet{dollasolomon2009}. 

The launch of the Hinode mission opened new possibilities, thanks to the EIS spectrometer. Recently, \citet{banerjee2009} have identified the signature of Alfv\'en waves in EUV spectra observed with Hinode/EIS spectrometer in a polar coronal hole through the increase of the nonthermal line-of-sight velocities up to $\simeq$ 150'' above the limb. These waves appears to propagate undamped up to $\sim 0.15-0.20$ R$_\odot$ above the limb, while it is at the moment unclear whether energy deposition in the corona occurs above this altitude. Similar work has been performed recently by \citet{gupta2010}, using spectroscopic observations of SOHO/SUMER and Hinode/EIS, where they conclude that the observed propagating disturbances are magneto-acoustic waves. Nevertheless, none of these works studied with EIS the behaviour of these waves at larger altitudes.

Hence, in order to investigate the energy deposition by Alfv\'en waves in polar regions above $\simeq 0.20$ R$_\odot$, between February 24--25, 2009 a special off-limb study of a polar coronal hole has been performed with the Hinode/EIS spectrometer (HOP 103, EIS study 340). As we describe here, these unique dataset allowed for the first time to study with EIS the line profile broadening up to $\simeq 0.40$ R$_\odot$. The plan of the paper is as follows. In \S~2, the observations acquired for this study and the data reduction techniques are outlined. In \S~3, results of electron densities, non-thermal velocities and 
Alfv\'en waves energy flux are presented. A discussion of the observational results and a comparison with theoretical studies are taken up in \S~4, together with the final conclusions.

\section{Observations and data reduction}

\subsection{EIS Study description}

The Hinode EUV Imaging Spectrometer \citep[EIS; ][]{culhane2007} observes high-resolution spectra in two wavelength bands: 170-212 \AA\ (short wavelength band) and 246-292 \AA\ (long wavelength band). The instrument has 4 available entrance slits: two are available for imaging, with projected widths of 40'' and 266'', and two are dedicated to spectroscopy with projected widths of 1'' and 2'', oriented along the North--South direction. Pixel sizes on the detector are 1'' (spatial dimension) and 0.0223 \AA\ (spectral dimension). For single exposures the maximum field-of-view in Solar-Y is 512'' \citep[see][for a detailed description of the EIS instrument]{korendyke2006,culhane2007}.

The aim of the Hinode/EIS study $\#$ 340 was to measure the non-thermal velocities of various coronal ions above a polar coronal hole up to $\simeq 0.4$ R$_\odot$ above the limb. Hence, between February 24--25, 2009, the 512'' long EIS slit was centered above the South polar coronal hole, 202'' above the limb, covering a range of altitudes from $\simeq 0.05$ R$_\odot$ below the limb up to $\simeq 0.48$ R$_\odot$ above the limb (Fig.~\ref{fig01}). In order to maximize the signal-to-noise ratio and perform reliable gaussian line fittings up to  $\sim 0.4$ R$_\odot$, the instrument acquired $\simeq 21.6$ hours of continuous ``sit \& stare'' observations with the 1'' slit centered at X=0'' and Y=-1162'' (155 exposures, 500 s exposure time). These observations allowed us to derive the first EIS ``deep field spectra'' of EUV spectral lines in the off-limb polar coronal hole. Moreover, in order to have an idea of the structures (plumes or inter-plumes) present in the surrounding corona, two spatial rasters, each $\simeq 1.19$ hours long, have been acquired covering 80'' in the X direction with the 2'' slit (40 exposures, 100 s exposure time - hereafter ``context study''). The first ``context study'' was performed before the ``sit \& stare'' observations, the second one at the end of the ``sit \& stare'' observations. The field of views (FOVs) covered with the spatial and temporal rasters are shown in Fig.~\ref{fig01}.

Due to Hinode telemetry limitations, for this study we were able to select only 10 spectral panels (32 spectral bins per panel, equivalent to $\simeq 0.71$\AA). These panels are centered over the following lines: O~{\sc vi} $\lambda$184.00, Fe~{\sc xii} $\lambda$186.85, Ca~{\sc xvii} $\lambda$192.82, Ca~{\sc xiv} $\lambda$193.87, Fe~{\sc xii} $\lambda$195.12, Fe~{\sc xiii} $\lambda$202.04, He~{\sc ii} $\lambda$256.32, Fe~{\sc xiv} $\lambda$264.78, Fe~{\sc xiv} $\lambda$274.77, Fe~{\sc xv} $\lambda$284.16. Over the above spectral intervals, all the following Fe lines have been detected in this study with good statistics (i.e. with line peak intensity at least 2 times larger than the surrounding background): Fe~{\sc x} $\lambda$193.71, Fe~{\sc xii} $\lambda$186.88, Fe~{\sc xii} $\lambda$193.51, Fe~{\sc xii} $\lambda$195.12 and Fe~{\sc xiii} $\lambda$202.04.

\subsection{Data reduction and analysis}

Acquired data have been first calibrated with the standard routines provided by the \textit{SolarSoftware} Hinode/EIS package, aimed also at removing instrumental effects such as the tilt of the EIS slits relative to the detector column's orientation, ``hot'' and ``warm'' pixels \citep[by interpolating missing data points; see]{young2009b}, and the line centroid shifts related to the Hinode spacecraft Sun-syncronous orbit, occurring with a period of $\simeq 100$ minutes. Second, by employing the periodic wavelength shift provided by the standard calibration routines \citep[recently improved by the EIS Team; see][]{kamio2010,kamio2011}, each line profile has been resampled bin per bin by fractional amounts of spectral pixels, preserving the total number of counts accumulated over the line profile (Fig.~\ref{fig02}, left panels). Third, each identified spectral line has been fitted with a gaussian profile, in order to derive the evolution with time of the integrated emission at each altitude covered by the 512 slit spatial bins. Fourth, a further correction has been applied to the intensity versus time maps in order to remove a periodic pointing displacement by $\pm \simeq 6$'' observed in the Y direction, related to the spacecraft orbit and observed as an apparent periodic limb oscillation (Fig.~\ref{fig02}, right panels).

Resulting intensities do not show clear transit of plumes/interplume structures dragged by solar rotation in and out of the spectrometer FOV (Fig.~\ref{fig03}, right panel). This is in agreement with the global diffuse appearance of the off-limb EUV corona during our observations, as shown by the context EIS observations acquired before and after the ``sit \& stare'' observations (Fig.~\ref{fig03}, left panels), and also by EIT images (Fig.~\ref{fig01}). No clear plume or inter-plume structures are visible at that time, and the line intensity evolution simply shows a very slow decrease during our observations. We were then unable to distinguish between these different structures. In order to have a better statistics, in this analysis we have summed over time all data acquired during the ``sit \& stare'' study. Hence, physical parameters we will derive in the next sections have to be considered as representative of an average plume/interplume plasma. Moreover, at higher altitudes, summing over $\simeq 21.6$ hours of observations was not sufficient to have a good statistic on the line profiles. Hence, as also done by \citet{banerjee2009}, profiles at the southern off-limb edge of the slit were averaged over different spatial intervals, summing over a larger number of pixels at higher altitudes. An example of line profiles obtained at 3 different altitudes with different spatial averages (considered in the analysis) is shown in Fig.~\ref{fig04}: it is clear that the spectral lines are detected with a good statistics even at 0.4 R$_\odot$ above the limb.

\subsection{Stray light estimate}

The line profiles obtained in the Fe~{\sc xii} $\lambda$195 spectral window (194.785--195.477 \AA) by averaging over the whole observation interval are shown in Fig.~\ref{fig05} (top left panel) at 3 different altitudes. This plot shows that on-disk two more spectral lines are present, centered approximately at $\lambda$194.828~\AA\ and $\lambda$195.416~\AA, as we derived with a multi-gaussian fitting. Both these lines disappear quickly moving off-limb, as shown by the average profiles at 50'' and 150'' off-limb in this Figure. The Fe~{\sc xii} $\lambda$195.119 is observed in our spectra at $\lambda$195.139, hence there is a shift by $\Delta\lambda=$0.020~\AA\ with respect to reference wavelenght. This means that the corrected centroids of the unidentified lines are $\lambda$194.808 and $\lambda$195.396.

The identification of these two spectral lines is not straightforward: the latest version of the CHIANTI spectral code \citep[v.7.0;][]{dere2009} provides in this wavelenght interval only 4 observed spectral lines (Fe~{\sc xii} $\lambda$195.119-$\lambda$195.179, Ni~{\sc xvi} $\lambda$195.275 and Zn~{\sc xx} $\lambda$195.379) and 2 theoretical but unobserved lines (Fe~{\sc ix} $\lambda$195.250 and Fe~{\sc x} $\lambda$195.316). Hence, the CHIANTI code does not provide any possible identification for the $\lambda$194.808 and $\lambda$195.396 lines which we have observed. In the literature, \citet{brown2008} identified with EIS in this range the Fe~{\sc xiv} $\lambda$195.274 and Fe~{\sc x} $\lambda$195.395 lines (observed at the limb and 20'' off-limb), while \citet{landiyoung2009} identified a line observed at $\lambda$195.415 as the Fe~{\sc vii} $\lambda$195.391 line. Other possible lines close to the Fe~{\sc xii} $\lambda$195.12 have been reported by \citet{young2009} (see their Fig.~6), which identified the Fe~{\sc xiv} $\lambda$195.25, Fe~{\sc x} $\lambda$195.40 and Ni~{\sc xv} $\lambda$195.52; Ni~{\sc xiv} $\lambda$195.27 was also reported by \citet{young2009}. A line observed at $\lambda$195.39 was identified by \citet{delzanna2009} as the Fe~{\sc viii} $\lambda$195.39 line. Nevertheless, Fe~{\sc xiv} $\lambda$195.25, Ni~{\sc xv} $\lambda$195.52 and Fe~{\sc viii} $\lambda$195.39 were reported in Active Regions in these works. In particular, the Fe~{\sc xiv} and Ni~{\sc xv} lines have peak emission temperatures of $\simeq 2.0 \times 10^6$ K and $\simeq 2.5 \times 10^6$ K, respectively, and no significant emission is expected from these lines in a coronal hole below 0.4 R$_\odot$ off-limb, where SUMER and EIS data show electron temperatures tipically below $1.1 \times 10^6$ K \citep[][]{landi2008,hahn2010}.

Hence, there are at least three possible candidates for the identification of the line we observed at $\lambda$195.396, i.e. Fe~{\sc viii} $\lambda$195.388, Fe~{\sc x} $\lambda$195.395 and Fe~{\sc vii} $\lambda$195.391. The main difference between these lines is obviously the peak temperature of their emissivities: while the Fe~{\sc x} line emissivity peaks at $T = 1.3 \times 10^6$ K, hence at typical coronal temperatures, the Fe~{\sc vii} emissivity peaks at $ T = 4.0 \times 10^5$ K (as provided by CHIANTI code), hence it is mainly a transition region line. This means that it is possible to distinguish between these two lines by looking at their intensity profiles with altitude. In fact, plots in the top-right panel of Figure ~\ref{fig05} show that the intensities of Fe~{\sc xii} $\lambda$195.12, Fe~{\sc xiii} $\lambda$202.04 and Fe~{\sc x} $\lambda$193.71 coronal lines have very similar decays off-limb, while the intensity of the unidentified line at $\lambda$195.396 decays much faster (in particular much faster than the Fe~{\sc x} $\lambda$193.71 line) and is mainly unobserved above $\simeq 120$'' off-limb ($\simeq$ 0.13 R$_\odot$). This suggests us that the line we observe at $\lambda$195.396 is not a coronal line and could be identified as the Fe~{\sc vii} $\lambda$195.391 transition region line. Nevertheless, a contribution also from the Fe~{\sc viii} $\lambda$195.39 line cannot be totally excluded, in principle, because the contribution function of this line peaks at a temperature of $\simeq 6.3 \times 10^5$ K and has a broader temperature distribution, making possible some contribution from the $1.1 \times 10^6$ K coronal hole plasma, even if (as far as we know), this line has been never reported off-limb in this kind of region.

The identification of our unidentified line as Fe~{\sc vii} $\lambda$195.391 cannot be verified by comparing its intensity with those observed in other lines from the same multiplet, because no other Fe~{\sc vii} lines are present in our dataset. Given the observed Fe~{\sc x} $\lambda$193.71 intensity, we can at least estimate the expected contribution by the Fe~{\sc x} $\lambda$195.395 line to the observed unidentified line. As provided by the CHIANTI spectral code (v. 7.0), the Fe~{\sc x} $\lambda$195.39/$\lambda$193.71 line ratio is almost constant in the off-limb coronal hole density range ($10^7-10^9$ cm$^{-3}$) on the order of $\simeq 0.50$. From the observed Fe~{\sc x} $\lambda$193.71 intensity we conclude that the expected Fe~{\sc x} $\lambda$195.39 intensity should be smaller than 12\% and 43\% of the total intensity of the unidentified line, 5'' and 390'' above the limb, respectively. These values represent only upper limits, because have been computed before the correction for stray light, i.e. before removing the part of the radiation which is emitted on-disk and that, following optical paths not originally intended in the spectrometer, is finally detected off-limb. Hence we conclude that the Fe~{\sc x} $\lambda$195.39 coronal emission is expected to be much smaller than the intensity observed in the unidentified line. Unfortunately, we have no other Fe~{\sc viii} lines in our dataset, hence we were not able to perform the same estimate of the expected Fe~{\sc viii} $\lambda$195.39 contribution to the total intensity of our unidentified line.

In summary, taking into account that, as mentioned above, a) the intensity of the unidentified line decays with altitude much faster than other coronal lines, b) that coronal emissions from other lines observed in this range (such as Fe~{\sc xiv} and Ni~{\sc xv}) are expected to be negligible in a coronal hole below 0.4 R$_\odot$ off-limb, and c) that the Fe~{\sc x} $\lambda$195.39 coronal emission is expected to be much smaller than the intensity of the unidentified line, we conclude that the observed unidentified line mainly is not due to coronal emission, even if some contribution from Fe~{\sc viii} $\lambda$195.39 cannot be totally excluded. For these reasons, the line observed at $\lambda$195.396 has been identified here as the Fe~{\sc vii} $\lambda$195.391 transition region line. If this interpretation is correct, its off-limb intensity should decay with altitude much faster than what we observed, becoming unobservable $\sim 3-4$'' off-limb (corresponding to an altitude of $\sim 2-3\cdot 10^3$km). This implies that the intensity we observe above this altitude can be ascribed to stray light of the EIS instrument.

As recently estimated by the EIS Team \citep[][]{ugarte2010}, the stray light contribution for on-disk observations is on the order of 2\% of the total intensity. In order to estimate the off-limb stray light contribution, we employed the same technique recently applied by \citet[][]{hahn2011}: given the intensity $I_{FeVII}(disk)$ of the Fe~{\sc vii} $\lambda$195.39 line observed on-disk, we assumed that 2\% of this intensity corresponds to the total Fe~{\sc vii} brightness scattered off in the bins looking in the off-limb corona. So, the stray light intensity is simply $I_{FeVII}(stray) = 0.02 \cdot I_{FeVII}(disk)$. The ratio between the $I_{FeVII}(stray)$ and the Fe~{\sc vii} intensity observed off-limb at different altitudes $h$ provides the stray light percentage curve $I_{stray}(h)$, which is given by
\begin{equation}
I_{stray}(h) = 0.02 \cdot I_{FeVII}(disk) / I_{FeVII}(h)\,\,(\%).
\end{equation}
Notice that the stray light computed with this technique may be slightly underestimated, because the 2\% estimate made by \citet[][]{ugarte2010} corresponds to the ``minimum stray light level above the dark current''. In any case, this estimate was also recently validated by \citet{hahn2011} by checking the ratio between the Si~{\sc x} $\lambda$256.3/261.0 lines observed immediately off-limb after removing from the He~{\sc ii} $\lambda$256.3 - Si~{\sc vii} $\lambda$256.3 blend the 2\% of the average He~{\sc ii} disk intensity. The stray light curve resulting from the above equation is shown in the bottom right panel of Fig.~\ref{fig05} (dotted line): the stray light contribution increases with increasing off-limb altitudes up to $\simeq 62$ \% at an altitude of 0.4 R$_\odot$ off-limb. Even if, as explained above, the straylight values could be in principle underestimated, both because the 2\% value is the minimum on-disk stray light level, and because a contribution from the Fe~{\sc x} $\lambda$195.39 and Fe~{\sc viii} $\lambda$195.39 coronal emissions cannot be totally excluded, this curve is in very good agreement with stray light values recently derived by \citet[][]{hahn2011} at 0.10, 0.15 and 0.20 R$_\odot$ with other spectral lines (triangles in the bottom right panel of Fig.~\ref{fig05}). This suggests that the Fe~{\sc x} $\lambda$195.39 and Fe~{\sc viii} $\lambda$195.39 contributions to the Fe~{\sc vii} $\lambda$195.39 intrensity was negligible, at least up to 0.2 R$_\odot$. 

The stray light curve $I_{stray}(h)$ as a function of altitude (Fig.~\ref{fig05}, bottom right panel, dotted line) is well reproduced by the following function:
\begin{equation}
I_{stray}(h) = A \arctan \left[ B\,(h-C) \right] + m\,h + q\,\, \mbox{(}\%\mbox{)}
\end{equation}
with $h$ measured in R$_\odot$ from the limb ($h > 0.03$ R$_\odot$). By fitting the stray light curve with this function we obtained the following values for the above parameters: $A=0.214$ (\%), $B=19.5$ (1/R$_\odot$), $C=0.158$ (R$_\odot$), $m=0.153$ (\%/R$_\odot$), $q=0.269$ (\%). The resulting fitting curve (Fig.~\ref{fig05}, bottom right panel, solid line) has been applied for the stray light correction of the intensity profiles of all the spectral lines. In particular, Fig.~\ref{fig05} (bottom left panel) shows the resulting off-limb stray light contribution (dashed line) to the observed Fe~{\sc xii} $\lambda$195.12 intensity (solid line), together with the intensity profile corrected for stray light contribution (dotted line). Notice that the intensity profiles before the stray light subtraction (Fig.~\ref{fig05}, top right panel) are quite flat above $\simeq 0.3$ R$_\odot$, suggesting possible stray light contamination, while after the stray light subtraction (Fig.~\ref{fig05}, bottom left panel) the Fe~{\sc xii} $\lambda$195.12 intensity profile (dotted line) turns out to be slowly decaying even above that altitude.

\section{Results}

The main target of this work is the estimate of the total energy flux $F_w$ transported by Alfv\'en waves off-limb in polar coronal hole up to $\simeq 0.4$ R$_\odot$. To this end, we need both a determination of the electron density $n_e$ and the non-thermal velocity $\xi$ at different altitudes. The determination of these three quantities ($n_e$, $\xi$, and $F_w$) is decribed in the next Sections.

\subsection{Electron densities}

As mentioned above, over the available spectral intervals, we have detected and identified the Fe~{\sc x} $\lambda$193.71, Fe~{\sc xii} $\lambda$195.12, Fe~{\sc xii} $\lambda$193.51, Fe~{\sc xii} $\lambda$186.88, and Fe~{\sc xiii} $\lambda$202.04 spectral lines. With these spectral lines, the better available density diagnostic technique is provided by the Fe~{\sc xii} $\lambda$195.12/$\lambda$186.88 density sensitive ratio \citep[see][]{young2007}. Electron densities obtained with this ratio are shown in Fig.~\ref{fig06} (left panel, open circles) up to an altitude of $\simeq 0.12$ R$_\odot$. These densities are in good agreement with those obtained from previous EIS off-limb observations \citep[e.g.]{banerjee2009} for a polar coronal hole (Fig.~\ref{fig06}, right panel, boxes and triangles). Nevertheless, above the altitude of $\simeq 0.12$ R$_\odot$ the intensity of the Fe~{\sc xii} $\lambda$186.88 starts to decay slower than the Fe~{\sc xii} $\lambda$195.12 intensity, resulting in a decrease of the $\lambda$195.12/$\lambda$186.88 ratio, and finally an unrealistic increase of derived densities. The reason for this behaviour could reside in the blending of the Fe~{\sc xii} $\lambda$186.88 line (which is also a blend between two Fe~{\sc xii} lines at $\lambda$186.854 and $\lambda$186.887) with the nearby S~{\sc xi} $\lambda$186.84 line, even if, as pointed out by \citet{young2009}, the contribution of the S~{\sc xi} line to the Fe~{\sc xii} line should be negligible.

In any case, for the above reason, we were unable to derive reliable electron density measurements above that altitude with the ratio $\lambda$195.12/$\lambda$186.88. Unfortunately the other well detected spectral lines do not provide any other good density sensitive ratio. Hence, in this work we derived the electron densities above the altitude of $\simeq 0.12$ R$_\odot$ with another technique. For any spectral line, the CHIANTI spectral code allows the computation of the contribution function $G_{line}(T_e,n_e,A_X)$ (erg cm$^3$ s$^{-1}$ sr$^{-1}$), which depends on the unknown plasma electron temperature $T_e$, density $n_e$, and also on the unknown elemental abundance $A_X$. Given $G_{line}$, the observed intensity $I_{line}$ of a spectral line formed by only collisional excitation can be written as
\begin{equation}
I_{line} = \frac{h \nu_{line}}{4\pi} \int_{LOS} \,G_{line}(T_e,n_e,A_X)\,n_e^2\,dz\,\,\, \mbox{(erg}\,\mbox{cm}^{-2} \mbox{s}^{-1} \mbox{sr}^{-1}\mbox{)}
\end{equation}
where the integration is performed along the line of sight (LOS). For spectral lines considered in this work the contribution from radiative excitation has been neglected, as usually done in the so-called ``coronal model approximation'' \citep[e.g.][]{mason1997}. The observed line intensity variations with altitude in the solar corona are due mainly to variations of $n_e$ and $T_e$, while the elemental abundance $A_X$ is usually assumed to be constant, in first approximation \citep[see e.g.][]{delzanna2002}. Hence, given a pair of possible values for the average density and temperature along the LOS ($<n_e>,<T_e>$), the expected line intensity $I_{line}$ can be estimated, by assuming a thickness along the LOS of the emitting plasma. By comparing the line intensity observed at a fixed altitude with all the expected values $I_{line}$ computed for a range of possible densities and temperatures, a curve on the ($<n_e>,<T_e>$) plane of allowed values can be derived. Hence, from a set of  intensities for different spectral lines observed at the same altitude, it is possible to perform a minimum chi-square analysis and the best ($<n_e>,<T_e>$) values have been derived by minimizing the difference between the observed and computed intensities for all the spectral lines (see e.g. Fig.~\ref{fig07}). 

In this work we have applied the above technique. In particular, for the computation we employed the intensities observed at different altitudes of the following iron spectral lines: Fe~{\sc xii} $\lambda$195.12, Fe~{\sc xii} $\lambda$193.51, Fe~{\sc xiii} $\lambda$202.04 and Fe~{\sc xii} $\lambda$186.88. These lines have been selected as those observed with the better signal-to-noise ratio up to an altitude of $\simeq 0.4$ R$_\odot$. From the comparison in Fig.~\ref{fig06} (left panel) it turns out that densities derived with our fitting technique are in quite good agreement with those derived from the ratio technique between $\simeq 0.03$ and $\simeq 0.12$ R$_\odot$, while at lower altitude are probably underestimated. This difference close to the limb is likely due to our assumption that the component due to the resonant scattering is negligible, while it has been shown that this hypothesis is not always correct \citep[see][]{schrijver2000}. Since the radiative and collisional components are roughtly proportional to $n_e$ and $n_e^2$, respectively, neglecting the radiative component may lead to an underestimate of the density. On the contrary, densities derived with the Fe~{\sc xii} line ratio are not affected by the background radiation field \citep[see discussion by][]{young2009}, even if densities derived with this technique can be not accurate close to the limb if the hypothesis of optically thin plasma falls down \citep[see][]{schrijver2000}. Also, Fig.~\ref{fig06} (right panel) shows that the densities values derived in this work are in good agreement with those derived by \citet{banerjee2009} with EIS data (triangles and boxes for plumes and inter-plumes, respectively), but slightly larger than densities derived by \citet{wilhelm2006} (upper and lower diamonds for plumes and inter-plumes, respectively) and by \citet{doyle1999} (dash-dotted line) with SUMER data.

Uncertainties shown in Fig.~\ref{fig06} have been computed from the chi-square curves (Fig.~\ref{fig07}), starting from statistical uncertainties on the selected spectral line intensities. The latters are very small, because data have been averaged over $\simeq 21.61$ hours of observation (i.e. smaller than 2\% at each altitude). A power law fitting curve to the densities we derived above  $\simeq 0.02$ R$_\odot$ is given by
\begin{equation}
n_e(h) = 5.644 \cdot 10^9 h^{-201.6} + 2.285 \cdot 10^8 h^{-10.42} + 5.656 \cdot 10^6 h^{-3.553} \, (\mbox{cm}^{-3})
\end{equation}
with $h$ (R$_\odot$) heliocentric distance measured from the Sun center. Because of the shape of the chi-square curves (see Fig.~\ref{fig07}), uncertainties in the resulting temperatures are larger than those on densities; hence we were not able to provide a reliable temperature profile at different altitudes. Resulting temperatures are on average $T = (1.3 \pm 0.3)\cdot 10^6$K. This temperature is in agreement with the fact that for the above analysis we employed only line emissions from Fe$^{11+}$ and Fe$^{12+}$ ions, whose maximum formation temperatures are around  $T \simeq 1.3 \cdot 10^6$K.

In order to verify the correctness of our results, the above densities and temperatures have been employed to compute again the expected line intensities: a comparison between the observed (diamond symbols) and computed (solid lines) intensities is shown in Fig.~\ref{fig08}. In the above computation we assumed the \citet{arnaudrothenflug1986} ionization equilibrium. Nevertheless, systematic uncertainties in the computation of expected intensities are mainly due to uncertainties in the atomic parameters and ionization equilibria assumed by the CHIANTI routine to compute the contribution functions, as also pointed out by \citet{delzannamason2005} for Fe~{\sc xii} lines. In particular, we estimate that different ionization equilibria correspond to a systematic uncertainty by 24\% and by 40\% for the Fe~{\sc xii} and Fe~{\sc xiii} lines, respectively. Corresponding upper and lower limits for the computed intensities are shown as dotted lines in Fig.~\ref{fig08}; this figure shows a general good agreement between the observed and computed line intensities at different altitudes, except for the Fe~{\sc xii} $\lambda$186.88 above $\simeq 0.12$ R$_\odot$, as already pointed out.

\subsection{Non-thermal velocities}

The non-thermal velocities of emitting ions have been determined from the observed line profiles as follows. The observed line profile $1/e$ width, $\sigma$, for any coronal spectral line can be written as
\begin{equation}
\sigma^2 = \frac{\lambda_0^2}{2c^2}\,\left( \frac{2k_B T_{ion}}{M_{ion}} + \xi^2 \right) + \sigma_I^2
\end{equation}
where $T_{ion}$ is the kinetic temperature of the emitting ion with mass $M_{ion}$, $\xi$ is the non-thermal broadening due to unresolved plasma velocities occurring along the line-of-sight and $\sigma_I$ is the instrumental line broadening. This formula is valid under the hypothesis that $\xi$ velocities have a Gaussian distribution. Given the observed line profile $\sigma$, the measured $\sigma_I$ and by assuming a value for $T_{ion}$, this relationship can be employed to estimate $\xi$. As recently pointed out by the EIS Team \citep[][]{young2011a}, the EIS instrumental width $\sigma_I$ is not constant with different $Y-$pixels along the slit. In this work we assumed $\sigma_I$ values provided by \citet[][]{young2011a}, which are on the order of $\sigma_I = 2.52$ pixels (i.e. $\sigma_I = 0.0563$ \AA) and $\sigma_I = 2.97$ pixels (i.e. $\sigma_I = 0.0663$ \AA) respectively at the bottom and the top of the EIS slit window for our observations (Fig.~\ref{fig01}). These values are slighly larger than the one derived by \citet{brown2008} ($\sigma_I = 2.42$ pixels $ = 0.054$ \AA) for the short wavelength band by assuming ion kinetic temperatures $T_{ion} = 10^6$ K and non-thermal mass motions $\xi = 30$ km s$^{-1}$. In any case notice that, given the uncertainties in the unknown quantities $T_{ion}$ and $\xi$ (see later) and the relatively small variation in $\sigma_I$ values along the slit given above, even in the case where $\sigma_I$ is over- or under-estimated, this will be mostly a systematic uncertainty, i.e. not affecting the relative variations of the derived $\xi$ values at different altitudes.

In this work we employed the Fe~{\sc xii} $\lambda$195.12 profiles, which give the better signal to noise ratio at all altitudes. As discussed by \citet[][]{young2009}, the Fe~{\sc xii} $\lambda$195 line is a blend between two Fe~{\sc xii} lines at $\lambda$195.119 and $\lambda$195.179, resulting in general in asymmetric Fe~{\sc xii} $\lambda$195 profiles when this line is observed on-disk. The line ratio Fe~{\sc xii} $\lambda$195.119/$\lambda$195.179 depends on the electron densities, as shown in Fig.~\ref{fig09} (left panel). In particular, for coronal hole density values we derived off-limb above a polar coronal hole ($n_e \sim 10^7-10^8$ cm$^{-3}$; see Fig.~\ref{fig06}), the $\lambda$195.179 line is expected to be $\sim 200-1000$ times fainter than the $\lambda$195.119 line. Hence, the Fe~{\sc xii} $\lambda$195 line is basically unaffected by the blend and the line profile can be safely fitted with a standard single gaussian profile off-limb above a polar coronal hole.

Nevertheless, the same is not true for the average Fe~{\sc xii} $\lambda$195 line emitted on-disk, which is clearly asymmetric (Fig.~\ref{fig09}, right panel). In particular, for electron density values measured on-disk at the base of the corona on the order of $10^9-10^{11}$ cm$^{-3}$ \citep[e.g.][]{young2009} the $\lambda$195.179 line is expected to be only $\sim 5-40$ times fainter than the $\lambda$195.119 line. By fitting the average on-disk profile (Fig.~\ref{fig09}, right panel, solid line) with two gaussians with the same width (i.e. the same Fe$^{11+}$ ion kinetic temperature) and separated by 0.06 \AA, it turns out that a $\lambda$195.119/$\lambda$195.179 line intensity ratio by 5.91 (corresponding to a density on the order of $\simeq 4 \times 10^{10}$ cm$^{-3}$) well reproduces the asymmetry of the Fe~{\sc xii} $\lambda$195 blend profile observed on-disk\footnote{See also \citet{young2011b} for a discussion of gaussian fitting and line blend for the EIS instrument.}.  

After a standard single gaussian fitting of the coronal line profiles, the $\sigma$ values have been corrected for the instrumental profile broadening, as explained above. Corrected $\sigma$ values provide an estimate for the broadening due at the same time to ion kinetic temperatures and plasma motions. In the literature, at least 2 different possible assumptions exist for the unknown quantity $T_{ion}$: the hypothesis of thermodynamic equilibrium ($T_{ion} = T_e$) and the hypothesis of ionization equilibrium ($T_{ion} = T_{max}$). Given the uncertainties mentioned above in the estimate of $T_e$, in this work we consider the ionization equilibrium hypothesis. By assuming \citep[as also done by previous authors, i.e.][]{mazzotta1998} that the unknown ion temperature equals the peak in the Fe~{\sc xii} $\lambda$195.12 line emissivity as provided by the CHIANTI spectral code ($T = 10^{6.1}$K), we estimate the broadening, $\xi$, due only to unresolved mass motions.

Nevertheless, another effect needs to be taken into account. As mentioned above, the Fe~{\sc xii} $\lambda$195.119-$\lambda$195.179 blend results in a broader Fe~{\sc xii} $\lambda$195 observed profile emitted from the disk. This implies that, for this line, opposite to what typically occurs for other spectral lines, the stray light contribution results in a broadening of the coronal line profiles, and not in a narrowing. In order to correct the Fe~{\sc xii} $\lambda$195.12 coronal profiles for the stray light broadening, we subtracted to the observed lines a broader stray light profile with a $\lambda$195.119/$\lambda$195.179 line intensity ratio by 5.91 and relative total intensity given by the stray light curve derived in the previous Section. In order to quantify the effect of the stray light profile subtraction, line widths have been also derived with a single gaussian fitting before this subtraction.

Resulting Fe~{\sc xii} line profile Full Widths at Half Maximum (FWHMs) and non thermal velocities $\xi$ (km s$^{-1}$) are shown in Fig.~\ref{fig10} (left and right panel, respectively), before and after the stray light subtraction (solid and dotted lines, respectively). Plotted FWHMs have been derived before the instrumental line broadening subtraction, while $\xi$ velocities have been computed by removing both the instrumental and thermal broadenings (formula 5). This figure shows that, within the error bars provided by the fitting procedure, a general trend is observed: after the stray light correction, the FWHM progressively increases up to an altitude of $\simeq 0.14$ R$_\odot$ above the limb, reaching a peak value of $\simeq 0.074$~\AA, then decreases at higher altitudes down to $\simeq 0.061$~\AA. Correspondingly, the estimated non-thermal velocity $\xi$ increases up to $\simeq 40$ km s$^{-1}$ at 0.14 R$_\odot$ above the limb, then decreases above that altitude down to $\simeq 10$ km s$^{-1}$ at 0.4 R$_\odot$ above the limb. The FWHM and $\xi$ profiles shown in Fig.~\ref{fig10} are quite well reproduced by the following second order polynomial fitting curves:
\begin{eqnarray}
FWHM(h) &=& 7.075 \cdot 10^{-2} + 4.424 \cdot 10^{-2} h - 1.720 \cdot 10^{-1} h^2\,\, \mbox{(\AA)} \nonumber \\
\xi(h) &=& 3.377 \cdot 10 + 8.638 \cdot 10 \, h - 3.565 \cdot 10^{2} h^2\,\, \mbox{(km/s)} \\
\end{eqnarray}
where $h$ (R$_\odot$) is the altitude above the limb; the above fitting curves are shown by dashed lines in the left and right panels of Fig.~\ref{fig10}.

\subsection{Alfv\'en waves energy flux}

In what follows we discuss the possible consequences of the working hypothesis that the unresolved plasma motions $\xi$ we observed are due to the propagation of Alfv\'en waves along open field lines above a polar coronal hole. Given the electron density and non-thermal velocity profiles derived above, it is possible to estimate the energy flux transfered into the corona by Alfv\'en waves. In particular, the Alfv\'en wave energy flux density $f_w$ (erg cm$^{-2}$ s$^{-1}$) for vertical propagation in a stratified, plane parallel atmosphere is given by \citep[e.g.][]{moran2001}:
\begin{equation}
f_w = \rho\, \xi^2 v_A = \sqrt{\frac{\rho}{4\pi}}\,\xi^2B
\end{equation}
where $v_A = B/\sqrt{4\pi\rho}$ is the Alfv\'en wave velocity, $B$ is the field strength, and $\rho$ is the mass density. This simplified relationship is correct only in the so called ``Wentzel-Kramers-Brillouin (WKB) approximation'', i.e. when the wavelengths are small compared to the background scale lengths. This approximation is valid on open fieldlines above $\sim 0.03-0.04$ R$_\odot$ off-limb, while non-WKB effects are important for wave propagation and reflection below this altitude \citep[see e.g.][]{hollweg1981,cranmerballe2005}, hence we can safely apply the above relationship for our data analysis. If the magnetic field flux $B \cdot A$ is conserved along opening flux tubes with cross sectional area $A$ carrying the  Alfv\'en waves, and if Alfv\'en waves propagate undamped, then the total Alfv\'en wave energy flux $F_w = f_w A$ (erg s$^{-1}$) should be a conserved quantity as the waves propagate outward.

Values of $f_w$ shown in Fig.~\ref{fig11} have been computed at different altitudes with the above formula by assuming a magnetic field $B=8$ G \citep[as done by][]{banerjee2009}. Results show a continuous decrease from $f_w \simeq 1.2\cdot 10^6$ erg cm$^{-2}$ s$^{-1}$ at 0.03 R$_\odot$ down to $f_w \simeq 8.5 \cdot 10^3$ erg cm$^{-2}$ s$^{-1}$ at 0.4 R$_\odot$. In this altitude range a linear fitting to the $f_w$ decay provides an Alfv\'en waves energy decay rate $\Delta f_w / \Delta h \simeq -4.5 \cdot 10^{-5}$ erg cm$^{-3}$ s$^{-1}$. Below 0.03 R$_\odot$, basically because of the much stronger density gradient, $f_w$ decays much faster with a rate $\Delta f_w / \Delta h \simeq -1.07 \cdot 10^{-3}$ erg cm$^{-3}$ s$^{-1}$. Implications from these results are discussed in the next section.

\section{Discussion and Conclusions}

In this work we have analyzed special long duration sit \& stare observations acquired by the Hinode/EIS spectrometer above the South polar coronal hole. The aim of this work is to study the EUV coronal line profile broadening observed off-limb and to provide a reliable measurement up to $\sim 0.4$ R$_\odot$ above the limb. In order to improve the signal-to-noise ratio, spectra acquired at different times have been averaged over the whole observation interval ($\sim 21.61$ hours), and a spatial average has been also performed for spectra acquired at higher altitudes. Results show that the Fe~{\sc xii} $\lambda$195 line profile (the strongest we detected) progressively broadens going from the limb up to $\sim 0.14$ R$_\odot$ above it, reaching a peak FWHM value of $\simeq 0.074$\AA. This result is in agreement with what was previously obtained with EIS by \citet{banerjee2009}, as shown in Fig.~\ref{fig12} (diamonds). 

\begin{figure}[th]
%\epsscale{.90}
\plotone{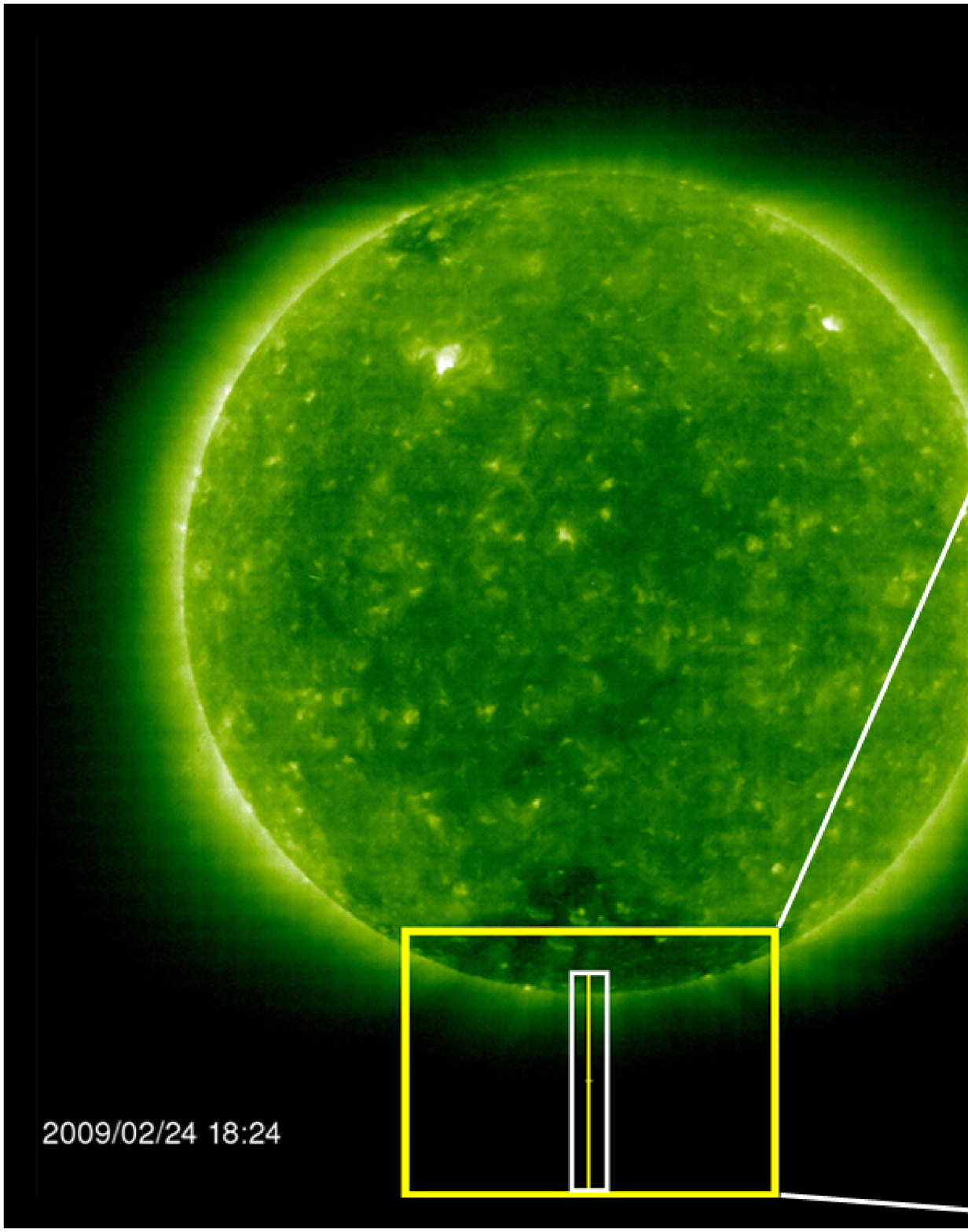}
\caption{Left: the EUV corona as seen by the SOHO/EIT telescope on February 24, 2009 with the Fe~{\sc xii} $\lambda$195 filter. The white box shows the Hinode/EIS spatial raster's FOV, while the yellow vertical line shows the slit position during the ``sit \& stare'' study. Right: a zoom on the region surrounded in the left panel by a yellow box. The EIS slit, extended off-limb out to $\sim 0.48$ R$_\odot$, is shown as yellow line and the spatial raster's FOV is illustrated by the white box.
} \label{fig01}
\end{figure}

\begin{figure}[th]
%\epsscale{.90}
\plotone{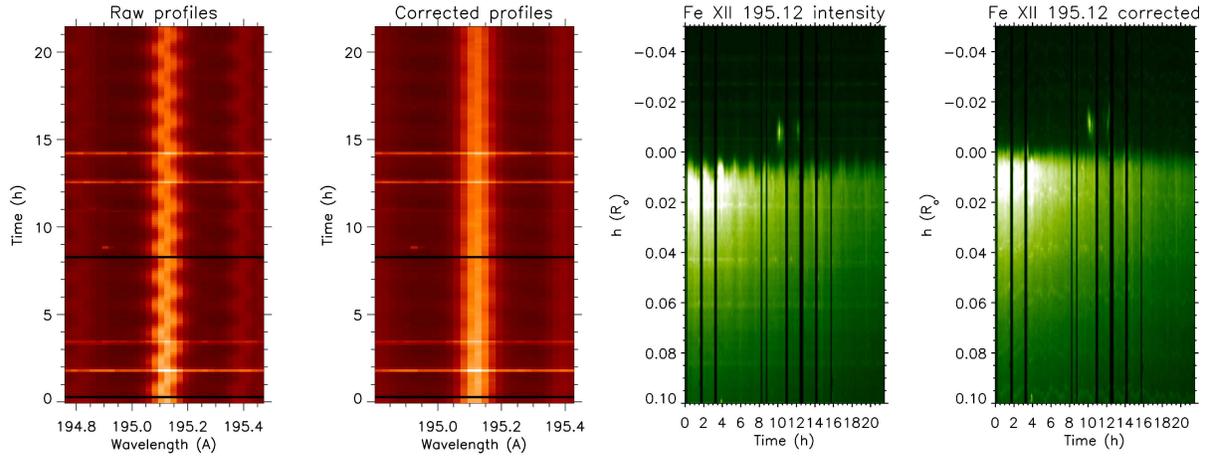}
\caption{Left: Fe~{\sc xii} $\lambda$195.12 line profiles ($x$-axis) averaged along the spatial direction (512'') as a function of time ($y-$axis) before (left) and after (right) the spectral resampling performed in order to remove the wavelength shift related to the Hinode spacecraft orbit. Right: a zoom over the Fe~{\sc xii} $\lambda$195.12 line intensities observed around the limb ($y-$axis) as a function of time ($x$-axis) before (left) and after (right) the spatial resampling performed in order to remove the pointing shift related to the Hinode spacecraft orbit.
} \label{fig02}
\end{figure}

\begin{figure}[th]
%\epsscale{.90}
\plotone{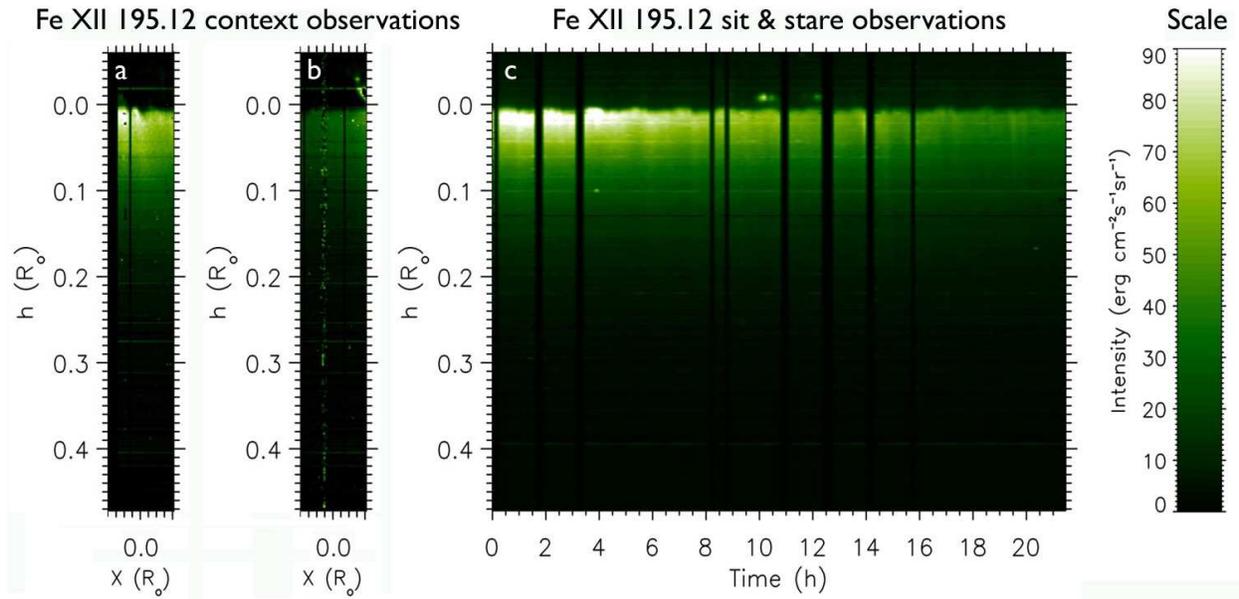}
\caption{The Fe~{\sc xii} $\lambda$195.12 line intensity as observed during the context and the sit \& stare observations. Distances are measured from the solar limb. Left panels (a-b): the line intensity as observed with two spatial rasters each $\sim$ 1 hour long, covering 80'' in the X direction with the 2'' slit, acquired before (a) and after (b) the sit \& stare observation.  Right panel (c): the line intensity evolution during the $\sim 21.6$ hours of ``sit \& stare'' observations with the 1'' slit.
} \label{fig03}
\end{figure}

\begin{figure}[th]
%\epsscale{.90}
\plotone{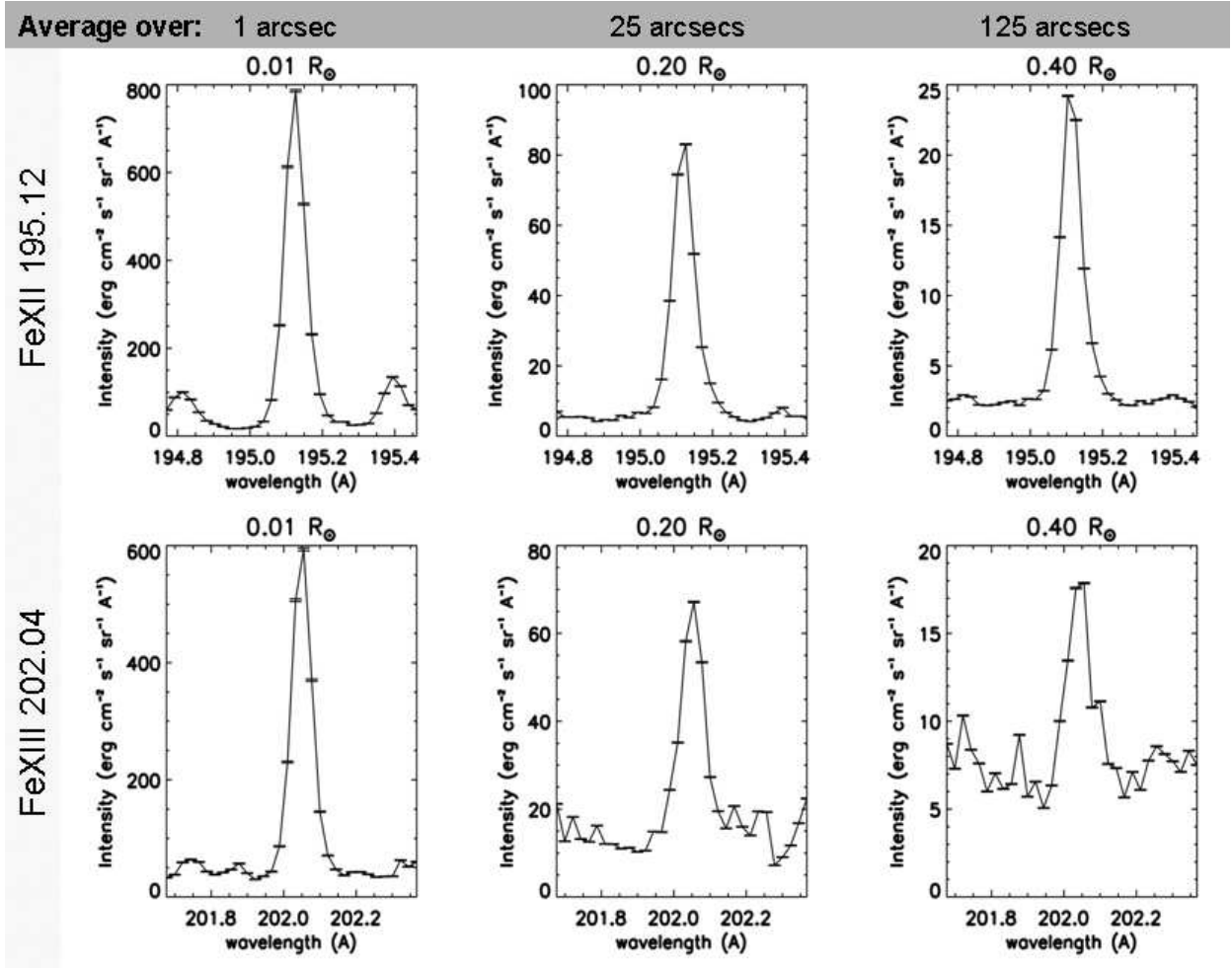}
\caption{Example of line profiles for the Fe~{\sc xii} $\lambda$195.12 (top) and Fe~{\sc xiii} $\lambda$202.04 (bottom) spectral lines at 3 different altitudes above the limb, as indicated at the top of each plot. Line profiles shown here were obtained by averaging line profiles over the whole sit \& stare observation interval and by further averaging over 125 (right column) and 25 (middle column) arcsecs along the slit.
} \label{fig04}
\end{figure}

\begin{figure}[th]
%\epsscale{.90}
\plotone{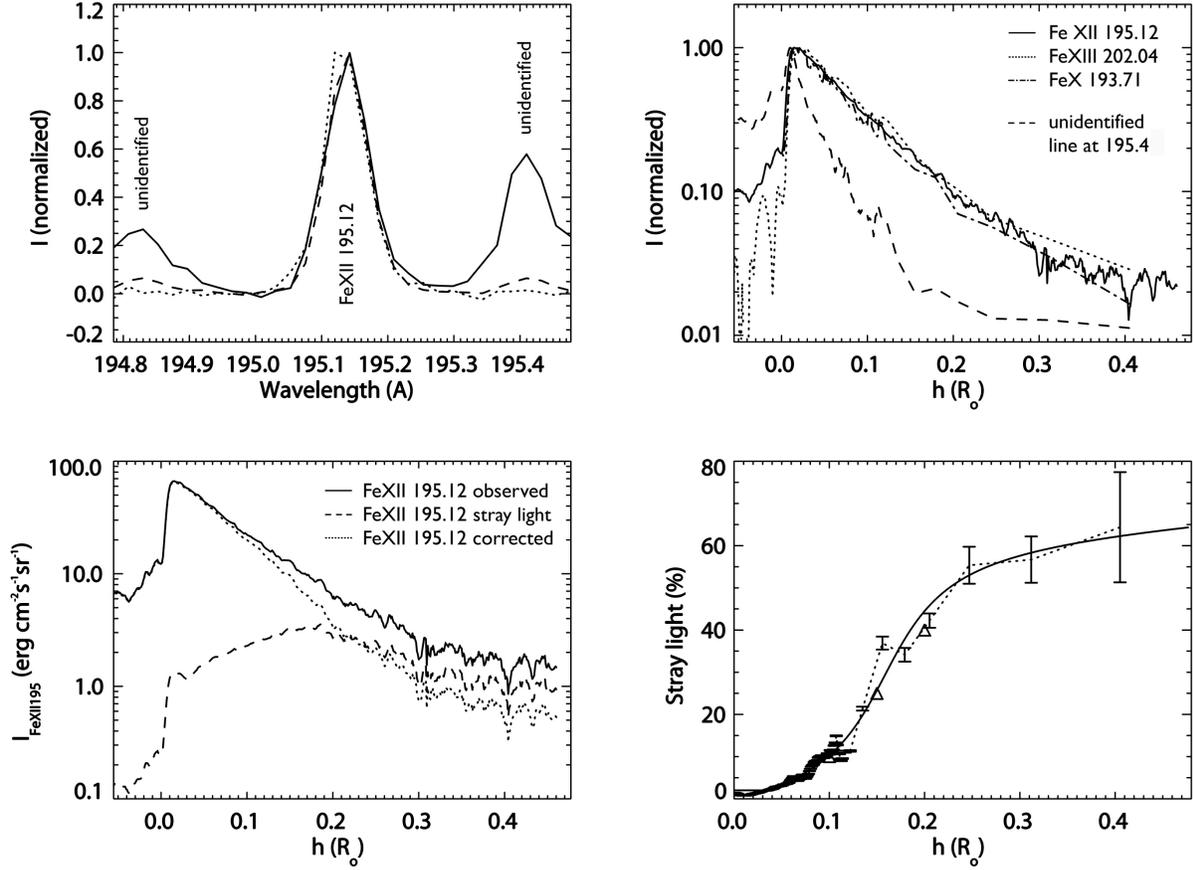}
\caption{Top left: the spectral profiles observed in the Fe~{\sc xii} $\lambda$195.12 panel 50'' on-disk (solid line), 50'' off-limb (dashed line) and 150'' off-limb (dotted line) from the solar limb. The solid line profile shows the Fe~{\sc xii} $\lambda$195.12 spectral line and 2 other lines, centered approximately at $\lambda$195.41 and $\lambda$194.83 (see text for discussion on the identification of these lines). Top right: normalized intensity profiles with altitude (logarithmic scale) for the Fe~{\sc xii} $\lambda$195.12 (solid), Fe~{\sc xiii} $\lambda$202.04 (dotted), Fe~{\sc x} $\lambda$193.71 (dash-dots) lines and for the unidentified line at $\lambda$195.41 (dashed). Bottom left: intensity profile with altitude (logarithmic scale) for the observed Fe~{\sc xii} $\lambda$195.12 line (solid), the fraction of Fe~{\sc xii} emission from the disk (dashed), and the Fe~{\sc xii} intensity corrected for stray light (dotted). Bottom right: the EIS stray light fraction estimated with the intensity of the $\lambda$195.41 line, identified as Fe~{\sc vii} $\lambda$195.415 (dotted line), compared with the values obtained by \citet[][]{hahn2011} (triangles). Solid line shows the fitting curve (eq.~2) of the derived values.
} \label{fig05}
\end{figure}

\begin{figure}[th]
%\epsscale{.90}
\plotone{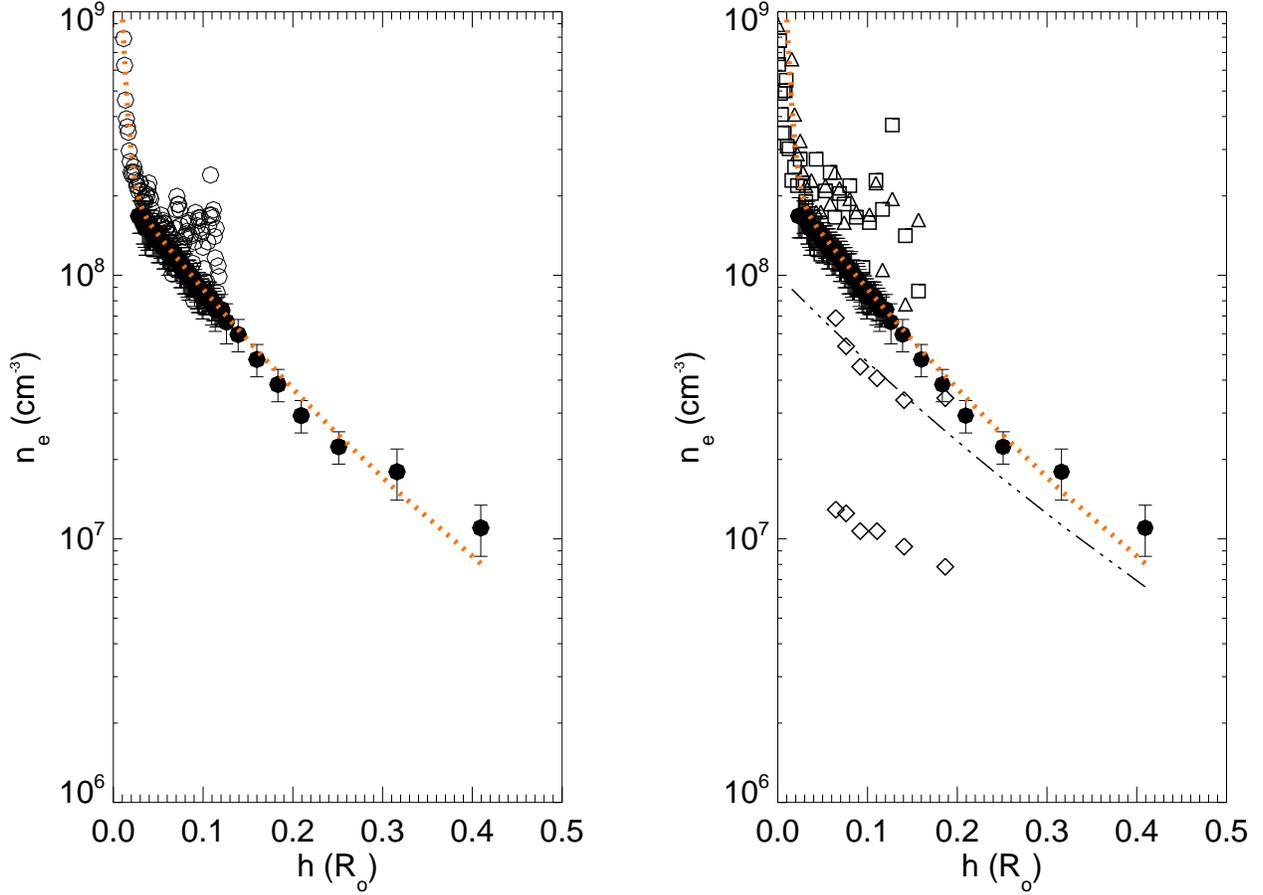}
\caption{Left: comparison between electron densities derived from the emission measure fitting procedure (filled circles) and those derived from the Fe~{\sc xii} $\lambda$195.12/$\lambda$186.88 line ratio (open circles). Dotted line shows the power law fitting curve (eq.~4). Right: electron densities derived with the emission measure fitting procedure (filled circles) and the power law fitting curve (dotted line). Density profiles derived by \citet{doyle1999} (dash-dots), \citet{wilhelm2006} (upper and lower diamonds for plumes and inter-plumes, respectively) and \citet{banerjee2009} (triangles and boxes for plumes and inter-plumes, respectively) are also shown for comparison.
} \label{fig06}
\end{figure}

\begin{figure}[th]
%\epsscale{.90}
\plotone{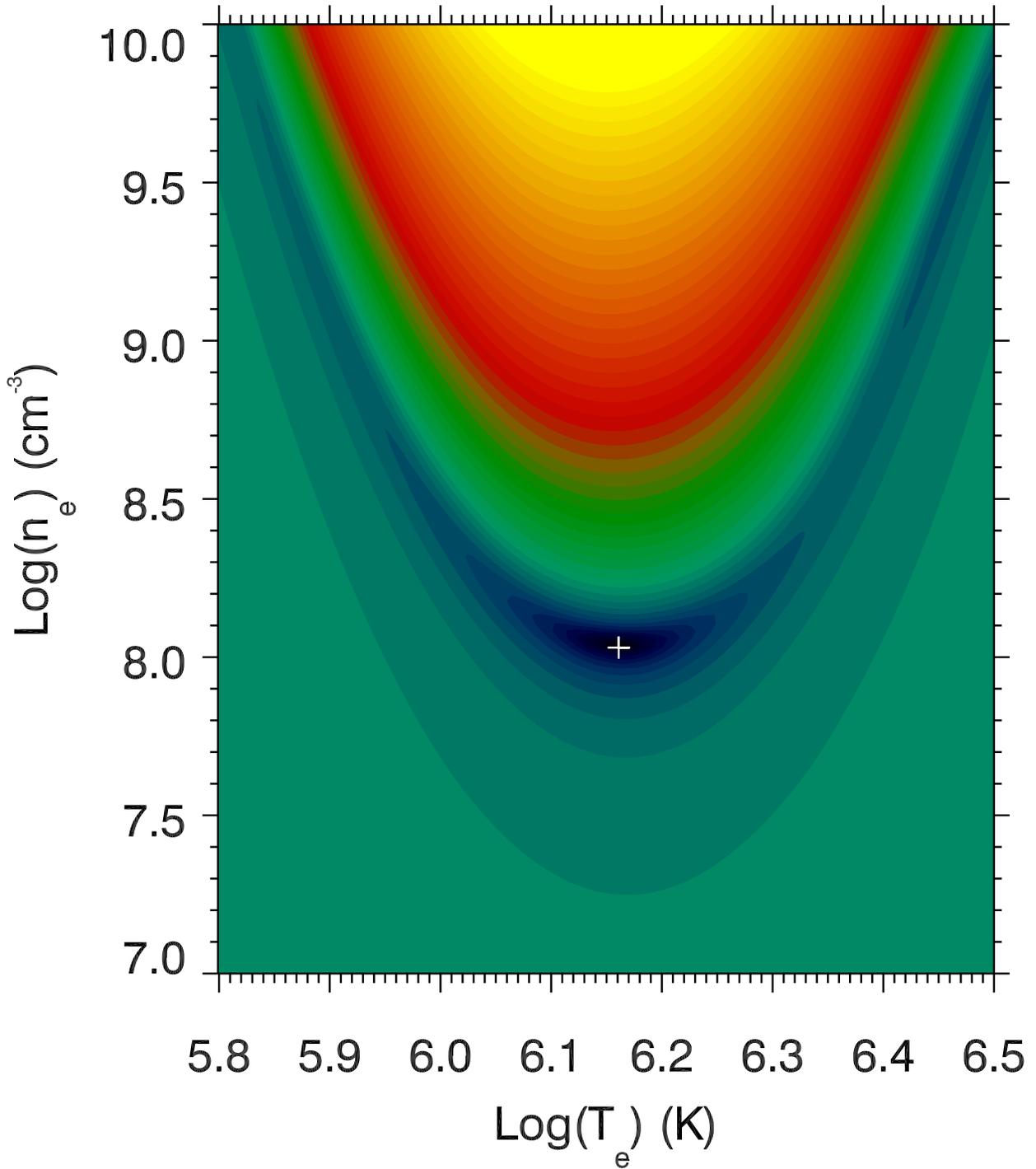}
\caption{Example of the chi-square distribution over the electron density vs. electron temperature plane for the line intensities observed at an altitude of 0.1 R$_\odot$ above the limb; solution point at that altitude is shown by the white cross.
} \label{fig07}
\end{figure}

\begin{figure}[th]
%\epsscale{.90}
\plotone{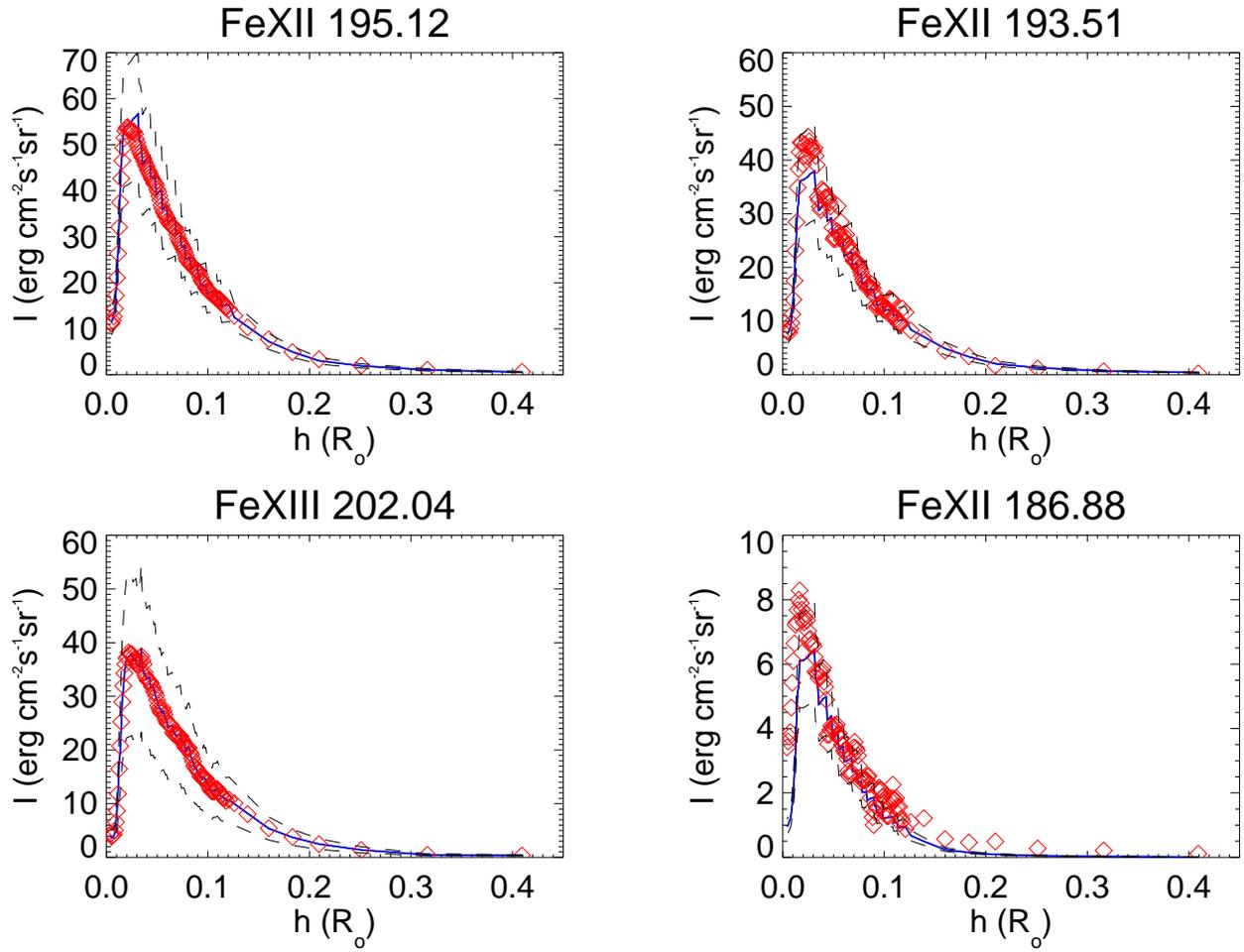}
\caption{Comparison between the observed line intensities (diamonds) and those computed (solid line) with densities and temperatures derived from the emission measure fitting procedure; upper and lower limits are shown as dotted lines (see text). 
} \label{fig08}
\end{figure}

\begin{figure}[th]
%\epsscale{.90}
\plotone{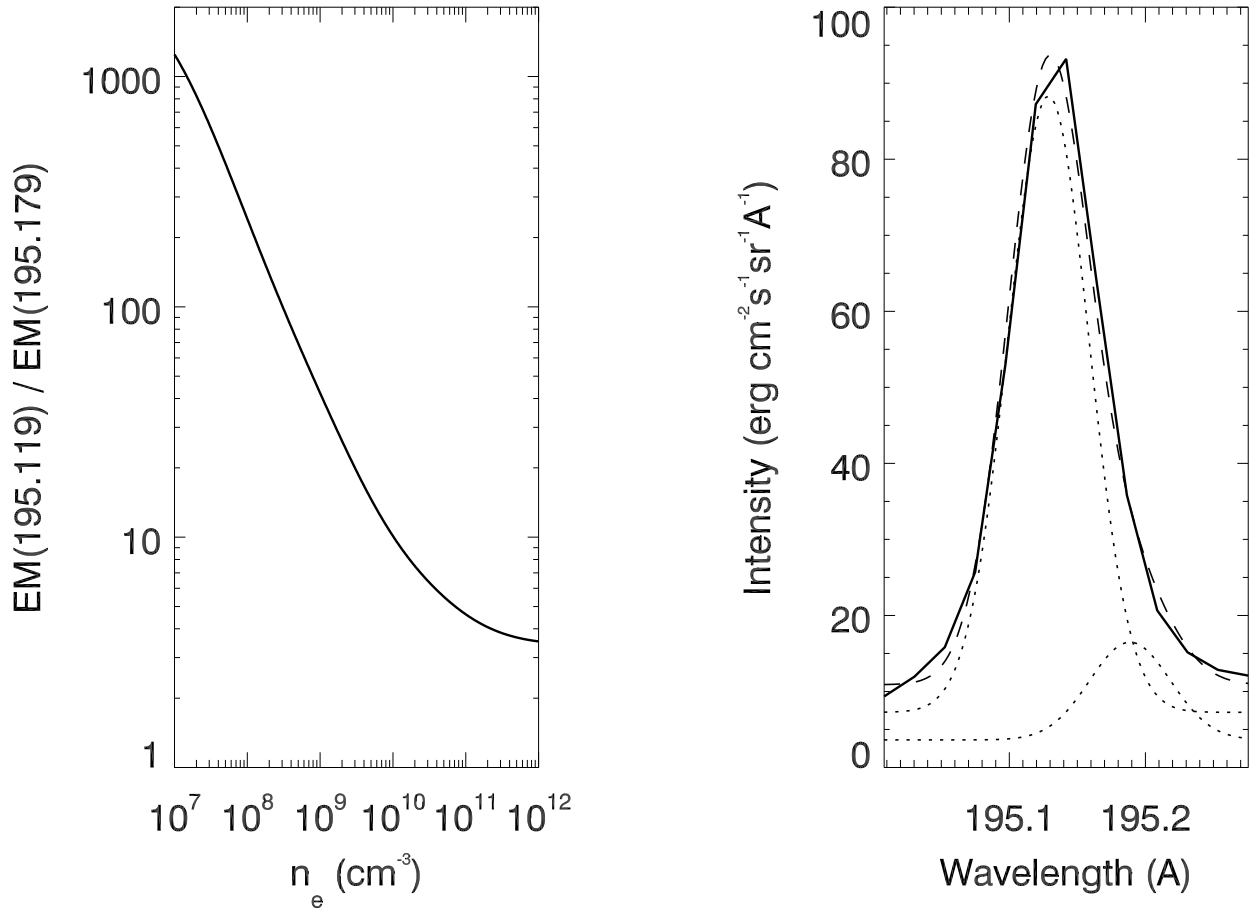}
\caption{Left: ratio between the emission measures ($y-$axis) of the Fe~{\sc xii} $\lambda$195.119 and $\lambda$195.179 lines provided by the CHIANTI spectral code (v.7.0) as a function of densities ($x-$axis). Right: Fe~{\sc xii} $\lambda$195.119-$\lambda$195.179 blend profile observed on-disk (solid), the corresponding two gaussians fitting curve (dashed), and single gaussians reproducing each spectral line (dotted).
} \label{fig09}
\end{figure}

Thanks to the FOV of the EIS off-limb study we performed (Fig.~\ref{fig01}), we were able also to study the FWHM versus altitude profile up to further distances than those ever explored by previous authors with the same instrument. Our results show that above 0.14 R$_\odot$ the FWHM progressively decreases down to $\simeq 0.061$~\AA~at $0.4$ R$_\odot$ above the limb. As we discussed in Section 2.3, the observed line width decrease above $\sim 0.14$ R$_\odot$ is not related to the EIS stray light contribution \citep[as reported by][for SUMER data]{dollasolomon2008,dollasolomon2009}, which has been estimated and removed from the observed Fe~{\sc xii} $\lambda$195 line profiles. Hence, as a first result, we exclude that the line width decrease observed above $\sim 0.14$ R$_\odot$ with EIS is an instrumental effect, and it needs to be physically explained. The main problem is that, as it is well known, the observed FWHM altitude profile is related to the altitude variation of two physical quantities (see formula 5), the ion kinetic temperatures $T_{ion}$ and non-thermal velocities $\xi$ (after correction for the instrumental line broadening $\sigma_I$ along the EIS slit), that can not be measured independently at the same time, unless some assumptions are made. Hence, derived non-thermal velocities are necessarily model-dependent. 

In the literature, many different assumptions have been made, hence providing in turn many different results. For instance: \citet{banerjee1998} assumed a constant $T_{ion} = 1 \cdot 10^6$K with altitude, in order to derive the $\xi$ profile for Si$^{7+}$ ions; \citet{tu1998} assumed a constant $\xi$ with altitude, different for different ions, in order to derive $T_{ion}$ as a function of the charge-to-mass ratios; \citet{moran2003} concluded that coronal ions have not a common velocity, hence non-thermal broadenings are not related to Alfv\'en waves, but they assumed a common $T_{ion}$. In general, if a common temperature is assumed for different ions, there is no common non-thermal speed, and vice-versa. In this work we were able to study the line width variation with altitude up to $\sim$ 0.4 R$_\odot$ only for the Fe~{\sc xii} $\lambda$195 line, while other spectral line profiles have a much lower signal to noise ratio. Non-thermal velocity profiles have been derived by assuming a constant $T_{ion} = 10^{6.1}$ at different altitudes, i.e. equal to the Fe~{\sc xii} $\lambda$195 line maximum emissivity, as provided by the CHIANTI spectral code. After a correction for the instrumental line broadening, this temperature provides us a progressive non-thermal velocity $\xi$ increase with altitude above the limb up to $\simeq 40$ km s$^{-1}$ at 0.14 R$_\odot$, then a decrease above that altitude down to $\simeq 10$ km s$^{-1}$ at 0.4 R$_\odot$.

\begin{figure}[th]
%\epsscale{.90}
\plotone{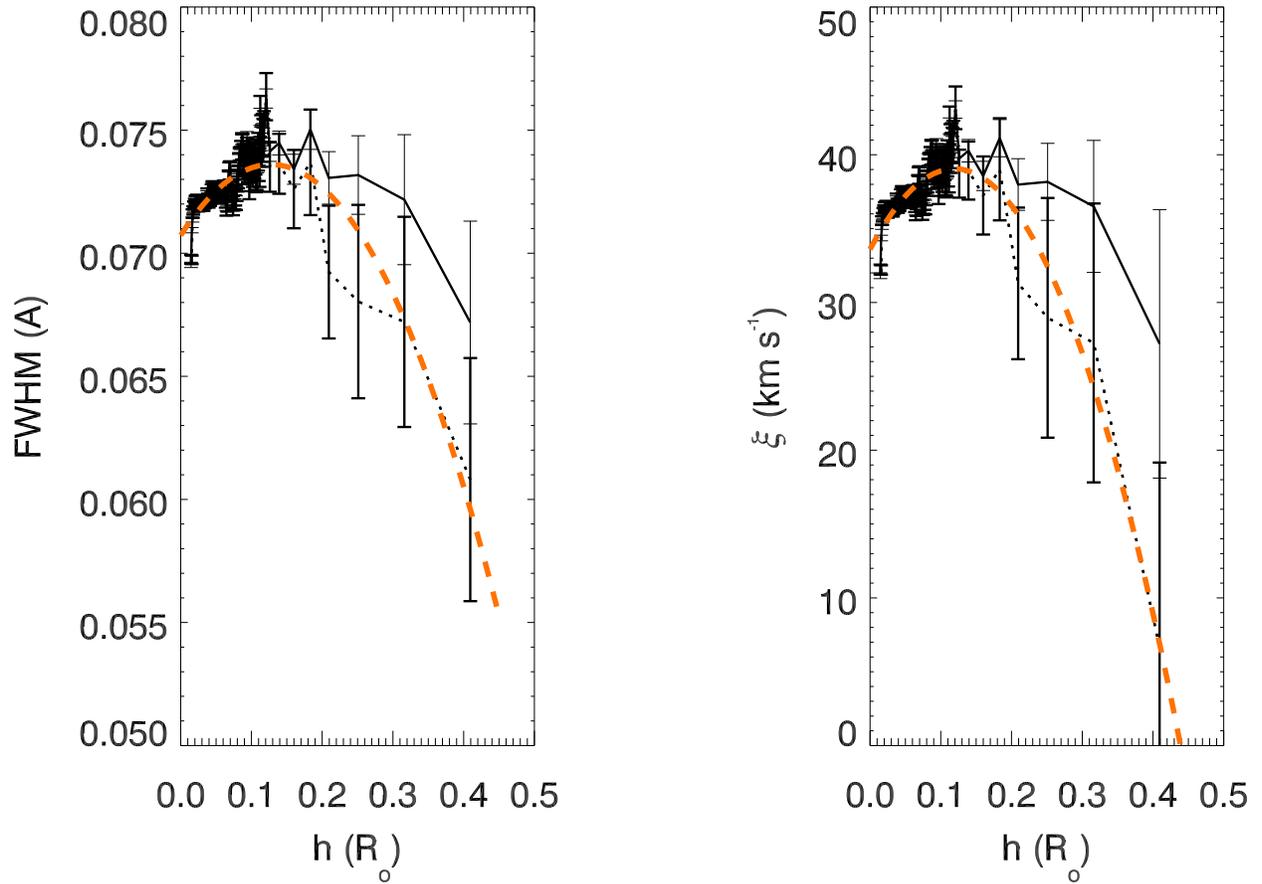}
\caption{Left: altitude profile of Fe~{\sc xii} $\lambda$195.12 FWHMs without correction for instrumental broadening as derived with a single gaussian fitting before (solid line) and after (dotted line) subtraction of the Fe~{\sc xii} stray light profile. Right: corresponding non-thermal velocities $\xi$ corrected for instrumental broadening computed before (solid line) and after (dotted line) subtraction of the Fe~{\sc xii} stray light profile. Dashed lines show parabolic fittings to the dotted curves (see text).
} \label{fig10}
\end{figure}

\begin{figure}[th]
%\epsscale{.45}
\plotone{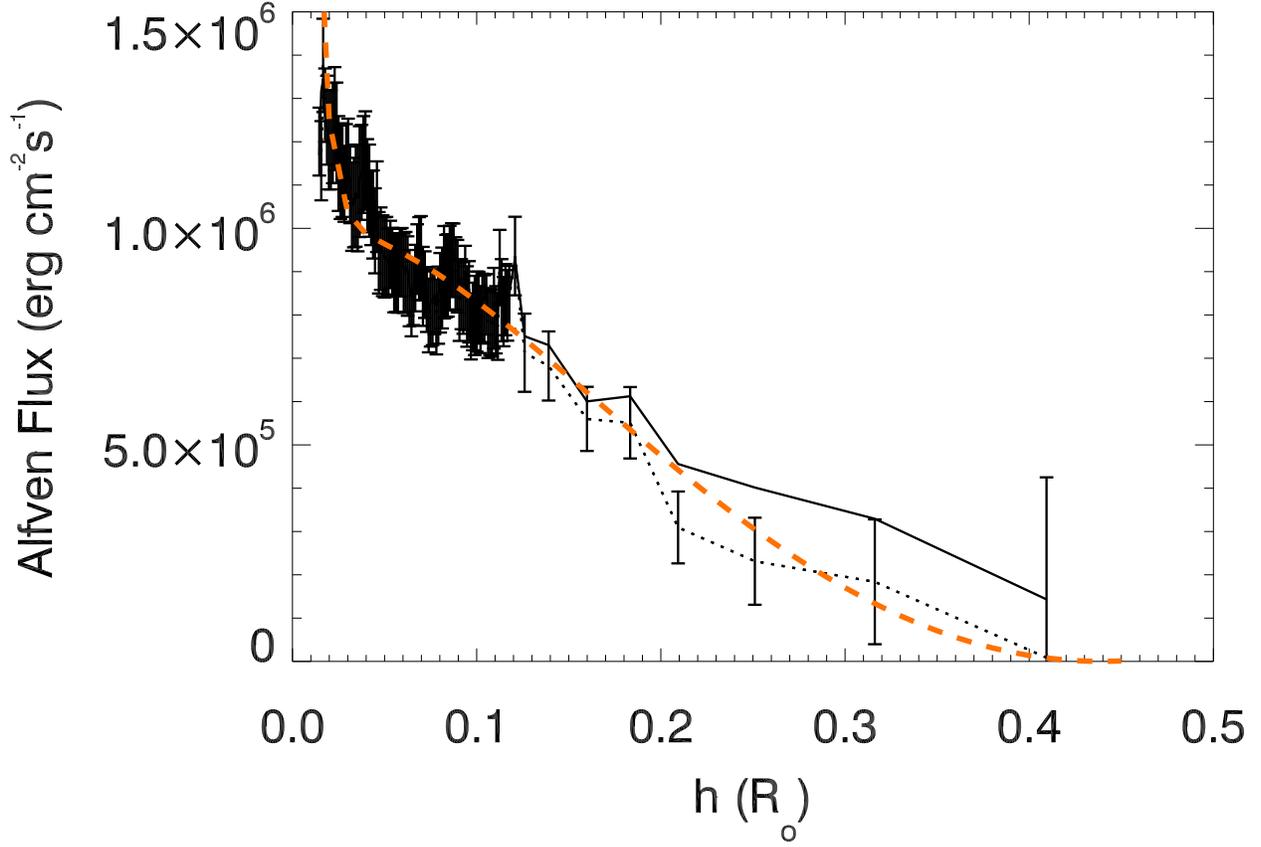}
\caption{The Alfv\'en wave energy flux $f_w$ as a function of altitude, computed by assuming magnetic flux conservation with non-thermal velocities $\xi$ derived before (solid line) and after (dotted line) subtraction of the stray light profile. This plot shows that the Alfv\'en wave energy flux progressively decays with altitude, with an average decay rate by $-4.5 \cdot 10^{-5}$ erg cm$^{-3}$ s$^{-1}$ above 0.03 R$_\odot$.
} \label{fig11}
\end{figure}

\begin{figure}[th]
%\epsscale{.45}
\plotone{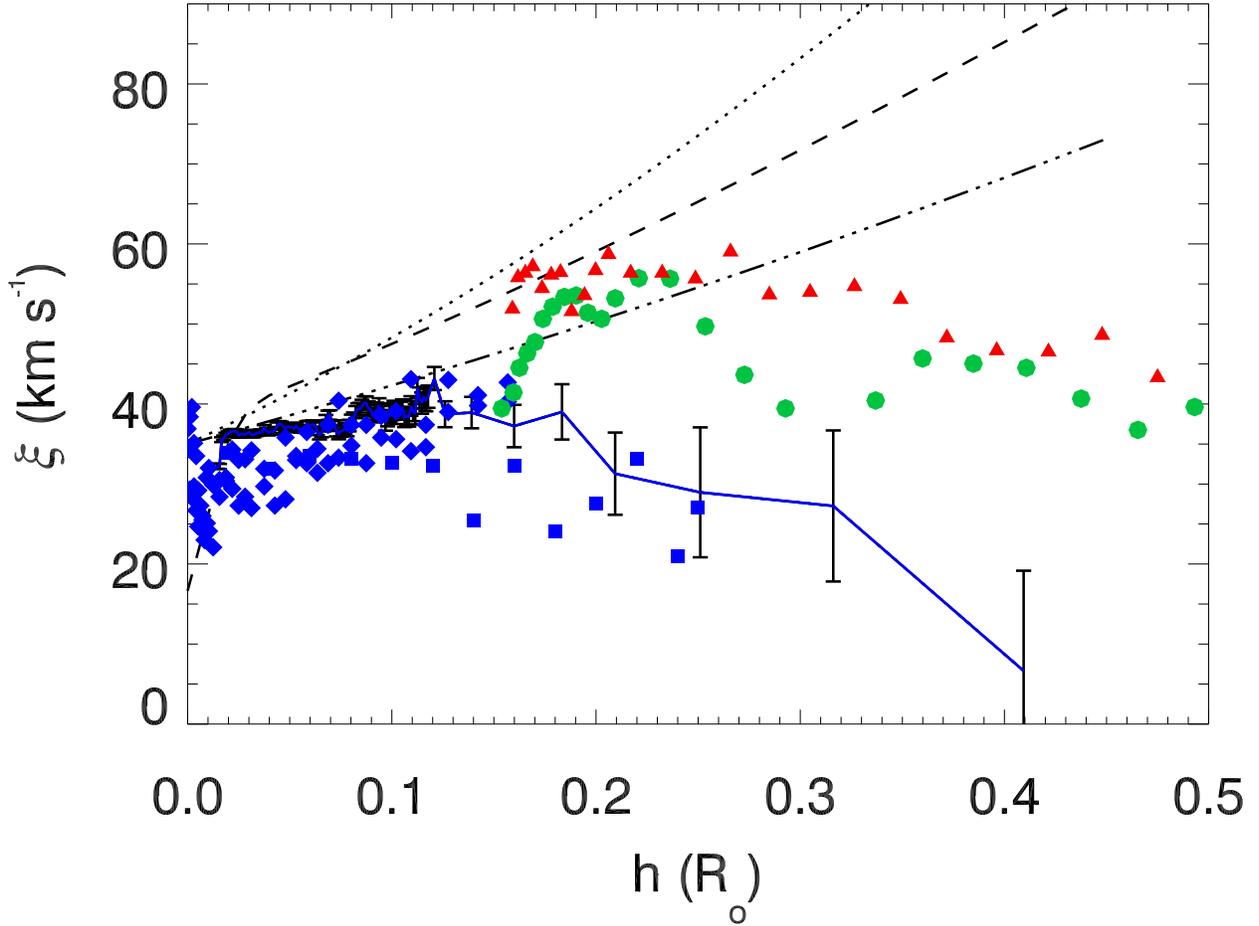}
\caption{Comparison between non-thermal velocities $\xi$ derived in this study (solid line) and those derived by several authors with different instruments above a polar coronal hole for the following Fe ions (in the on-line color version same colors corresponds to same ions): Fe$^{11+}$ derived with Hinode/EIS \citep[diamonds; ][]{banerjee2009}, Fe$^{11+}$ derived with SOHO/SUMER \citep[square boxes; ][]{moran2003}, Fe$^{12+}$ (triangles) and Fe$^{13+}$ (circles) derived with spectroscopic total eclipse observations \citep[][]{singh2011}. These values are compared with those expected for propagation of undamped Alfv\'en waves by assuming density profiles by \citet[][]{doyle1999} (dash-dotted line), \citet[][]{guhat1999} (dotted line) and the density profile derived in this work (dashed line).
} \label{fig12}
\end{figure}

In this work we also derived a coronal hole electron density profile for the off-limb corona up to 0.4 R$_\odot$ (eq.~4) with two methods: the line ratio technique and a minimum chi-square method aimed at reproducing the observed line intensity decays with altitude. The density and non-thermal velocity profiles derived led us to conclude that the Alfv\'en wave energy flux density $f_w$ continuously decreases with altitude above the limb from $f_w \simeq 1.2\cdot 10^6$ erg cm$^{-2}$ s$^{-1}$ at 0.03 R$_\odot$ down to $f_w \simeq 8.5 \cdot 10^3$ erg cm$^{-2}$ s$^{-1}$ at 0.4 R$_\odot$, with an average decay rate by $-4.5 \cdot 10^{-5}$ erg cm$^{-3}$ s$^{-1}$. Hence, the Alfv\'en waves energy flux density is not conserved (Fig.~\ref{fig11}), not even in the lower coronal hole region where the line profile FWHM increases with altitude (Fig.~\ref{fig10}). The disagreement with the behaviour expected for undamped Alfv\'en waves becomes more evident at larger altitudes. In fact, in the hypothesis of undamped Alfv\'en waves propagation, densities and non-thermal velocities should be related by $\xi \propto \rho^{-1/4}$, as already pointed out by \citet{hollweg1990,moran2001}. Hence, given a density profile $\rho(h)$, it is possible to compute the corresponding non-thermal velocity profile $\xi(h)$ that should be observed for undamped waves. Three example $\xi(h)$ curves are shown in Fig.~\ref{fig12} by assuming coronal hole density profiles provided by \citet[][]{doyle1999} (dash-dotted line), \citet[][]{guhat1999} (dotted line) and the density profile derived in this work (dashed line). A comparison between these curves and non-thermal velocities for Fe$^{11+}$ (diamonds), Fe$^{12+}$ (triangles) and Fe$^{13+}$ (circles) ions measured with different techniques in a coronal hole show that, even taking into account the broad scattering of different measurements (hence the large uncertainties), non-thermal velocity measurements could be in agreement with the propagation of undamped Alfv\'en waves up to no more than $\sim 0.2$ R$_\odot$, while at higher altitudes the disagreement with this hypothesis is evident.

The total Alfv\'en wave energy flux reported here is larger than what required to heat the quiet corona \citep[$3 \cdot 10^5$ erg cm$^{-2}$ s$^{-1}$;][]{withbroe1977}. Nevertheless, a physical explanation for this wave dissipation is needed. For instance, \citet{zaqarashvili2006} demonstrated that Alfv\'en and sound waves are decoupled for values of the plasma $\beta$ parameter $<1$, but become coupled when $\beta = 1$, allowing energy transfer from Alfv\'en to sound waves. If this energy is transferred to sound waves above the chromopsphere, then these waves will quickly dissipate their energy in the low density corona by inducing shock waves. This could be a possible explanation for the observed Alfv\'en wave damping: for instance following \citet{gary2001} the plasma $\beta$ above an active region approaches unity around 0.2 R$_\odot$, while outside active regions can approach unity even at lower heights. Alternative explanations have also been proposed. \citet{dwivedisrivastava2006} showed that, by taking into account the effect of viscosity and magnetic diffusivity in the radial propagation of Alfv\'en waves, an Alfv\'en wave energy flux density decrease is expected in polar coronal holes above 0.21 R$_\odot$; this decrease is more pronounced for low frequency ($10^3$ Hz) compared to high frequency ($10^4$ Hz) waves. This effect was confirmed by \citet{srivastavadwivedi2007} also for the more general case of oblique propagation. We notice here that the decay obtained by \citet{dwivedisrivastava2006} for the energy flux density of Alfv\'en waves is in qualitative agreement with our results (Fig.~\ref{fig11}) in the case of dissipation due to magnetic diffusivity alone for low frequency ($<10^3$ Hz) waves \citep[see][Fig.~2, middle panel]{dwivedisrivastava2006}. 

More recent observational results demonstrate the existence of ubiquitous low frequency ($10^{-2}-2 \cdot 10^{-3}$ Hz) Alfv\'en waves propagating throughout the quiescent atmosphere with sufficient energy to accelerate the solar wind and heat the corona \citep[][]{mcintosh2011} and suggest that low-frequency Alfv\'enic motions may suffer from significant damping as they propagate along magnetic structures \citep[][]{tomczyk2009}. Nevertheless, other theoretical works suggest that Alfv\'en waves should propagate undamped at these low altitudes above the limb. For instance, \citet{cranmerballe2005} developed a comprehensive model for the propagation of Alfv\'en waves in a coronal hole from the photosphere to the interplanetary medium, and demonstrate that the frequency-integrated velocity amplitude should increase monotonically with height from the chromosphere up to 10 R$_\odot$. In the present work we demonstrate that the FWHM decay of EUV lines (reported also by previous authors) cannot be simply ascribed to instrumental effects (i.e. stray light). In principle, alternative physical explanations for this observational effect, other than the damping of Alfv\'en waves, could be proposed. But a challenge remains to understand how to explain on the one hand the non-thermal velocity decay of heavy ions reported here in the EUV low coronal emission ($h < 0.4$ R$_\odot$), and on the other hand the very large non-thermal velocities of heavy ions discovered in the UV spectra emitted by the intermediate and extended corona \citep[$h > 0.5$ R$_\odot$; see review by][and references therein]{kohl2006}.

\acknowledgments
The authors would like to thank E. Landi, P. Young and G. Del Zanna for useful discussion on the line identification, and M. Hahn for useful suggestions on the EIS straylight estimate. We are also grateful to the anonymous Referee for very important comments on our work. We acknowledge support from ASI/INAF I/023/09/0 contract. Hinode is a Japanese mission developed and launched by ISAS/JAXA, with NAOJ, NASA and STFC (UK) as partners. It is operated by these agencies in co-operation with ESA and NSC (Norway). CHIANTI is a collaborative project involving researchers at NRL (USA) RAL (UK), and the Universities of: Cambridge (UK), George Mason (USA), and Florence (Italy).

\clearpage

\clearpage

\clearpage

\clearpage

\clearpage

\clearpage

\clearpage

\clearpage

\clearpage

\clearpage

\clearpage

\clearpage

\end{document}